\documentclass[12pt]{article}
 \usepackage[cp1251]{inputenc}
 \usepackage[english]{babel}
 \usepackage{amsmath, latexsym, amssymb, bm, array, graphics, eucal,
 amsfonts}
 \makeatletter
 \renewcommand{\@biblabel}[1]{#1.\hfill}

 \makeatother
 
\usepackage[dvips]{graphicx}

\begin{document}

 \thispagestyle{empty}
 \renewcommand{\abstractname}{\ }
 \large
 \renewcommand{\refname}{\begin{center} REFERENCES\end{center}}
\newcommand{\const}{\mathop{\rm const\, }}
 \begin{center}
\bf Friedel-Type Oscillations in the Problem of Skin Effect in
Degenerate Collisionless Plasma
\end{center}\medskip
\begin{center}
  \bf  A. V. Latyshev\footnote{$avlatyshev@mail.ru$} and
  A. A. Yushkanov\footnote{$yushkanov@inbox.ru$}
\end{center}\medskip

\begin{center}
{\it Faculty of Physics and Mathematics,\\ Moscow State Regional
University,  105005,\\ Moscow, Radio str., 10--A}
\end{center}\medskip

\begin{abstract}
It is shown that  a Friedel type   oscillations
accompany skin effect in degenerate plasma of a metal. It was learnt
earlier that Friedel oscillations take place under charge screening
in quantum plasma. However the nature of Friedel oscillations is not
in the quantum character of the plasma, but in the features of
degenerate Fermi distribution, namely, in its sharp transformation
into zero directly just the other side of the Fermi surface. This
circumstance leads to the Frieldel-type oscillations under anomalous
skin effect.

\medskip

electron gas

\end{abstract}

\begin{center}\bf
  1. Introduction
\end{center}

It is considered \cite{Landau}, that under penetration into
degenerate plasma the transversal electric field in the problem of
skin effect in the infrared area changes according to exponential
law
$$
\mathbf{E}=\mathbf{E_0}e^{-x/\delta}, \qquad
\delta=\dfrac{c}{\omega_p},
\eqno{(1)}
$$
where $c$ is the light speed, and $\omega_p$ is plasma (Langmuir)
frequency.

It is well known also (see, for instance, \cite{K}), that
dimensionless electric field in the problem of skin effect has the
following form
$$
\dfrac{E(x)}{E'(0)}=
\dfrac{al}{\pi}\int\limits_{-\infty}^{\infty}
\dfrac{e^{ik_1x_1}dk_1}{\varepsilon_{tr}(k_1)-ak_1^2}.
\eqno{(2)}
$$

Here $x_1$ is the dimensionless coordinate, $x_1=\dfrac{x}{l}$,
$k_1=kl$ is the dimensionless wave number, $k$ is the dimensional
wave number, $l=v_F\tau$ is the mean free path of electrons,
$$
a=\Big(\dfrac{c\varepsilon}{v_F\Omega}\Big)^2,\qquad
\Omega=\dfrac{\omega}{\omega_p}, \qquad
\varepsilon=\dfrac{\nu}{\omega_p},
$$
$\varepsilon_{tr}$  is the transversal dielectric permeability,
$$
\varepsilon_{tr}=1-\dfrac{3}{4\Omega (k_1\varepsilon)^3}\Bigg[
2(\Omega+i\varepsilon)(k_1\varepsilon)+\Big[(\Omega+i\varepsilon)^2-
(k_1\varepsilon)^2\Big]\ln\dfrac{\Omega+i\varepsilon-k_1\varepsilon}
{\Omega+i\varepsilon+k_1\varepsilon}\Bigg].
$$

In the present work it is shown that the dielectric permeability has
Kohn singularities  (see \cite{Kohn0} -- \cite{Kohn3}), which lead
to electric field  Friedel kind oscillations
\cite{friedel1} -- \cite{Grassme}.

\begin{center}
  \bf 2. Problem Solution
\end{center}

The quantity $\varepsilon_{tr}$ is regular under all values of the
frequency of the oscillations of the electric field and the wave
number. However, under the collision frequencies tending to zero,
i.e. under  $\varepsilon\to 0$ the wave number derivative
$\varepsilon_{tr}$ has singularities. These singularities are
analogous to so called Kohn singularities, which take place at
quantum longitudinal dielectric permittivity. It is known that Kohn
singularities result in change of the asymptotic form of the
electric field of the electric charge. Instead of the Debay
screening the slowly receding Friedel oscillations take place. On
the Fig. 1 the graph of the wave number derivative of the dielectric
permittivity is presented. We can easily see the features indicated
above on the graph. In accordance with this fact the change of the
asymptotic form of the electric field takes place under
skin effect.

On the Fig. 1  Kohn kind singularities for the
case of sodium are presented. We put the derivative
$d\varepsilon_{tr}/dq$ on the vertical axis. We consider two
cases of the frequencies $\Omega=0.08$ and $\Omega=0.1$.
Singularities of the derivative
$d\varepsilon_{tr}/dq$ in the points $q=0.08$ and $q=0.1$ is
seen on the graph.

\begin{figure}[h]
\begin{flushleft}
\includegraphics[width=18.0cm, height=14cm]{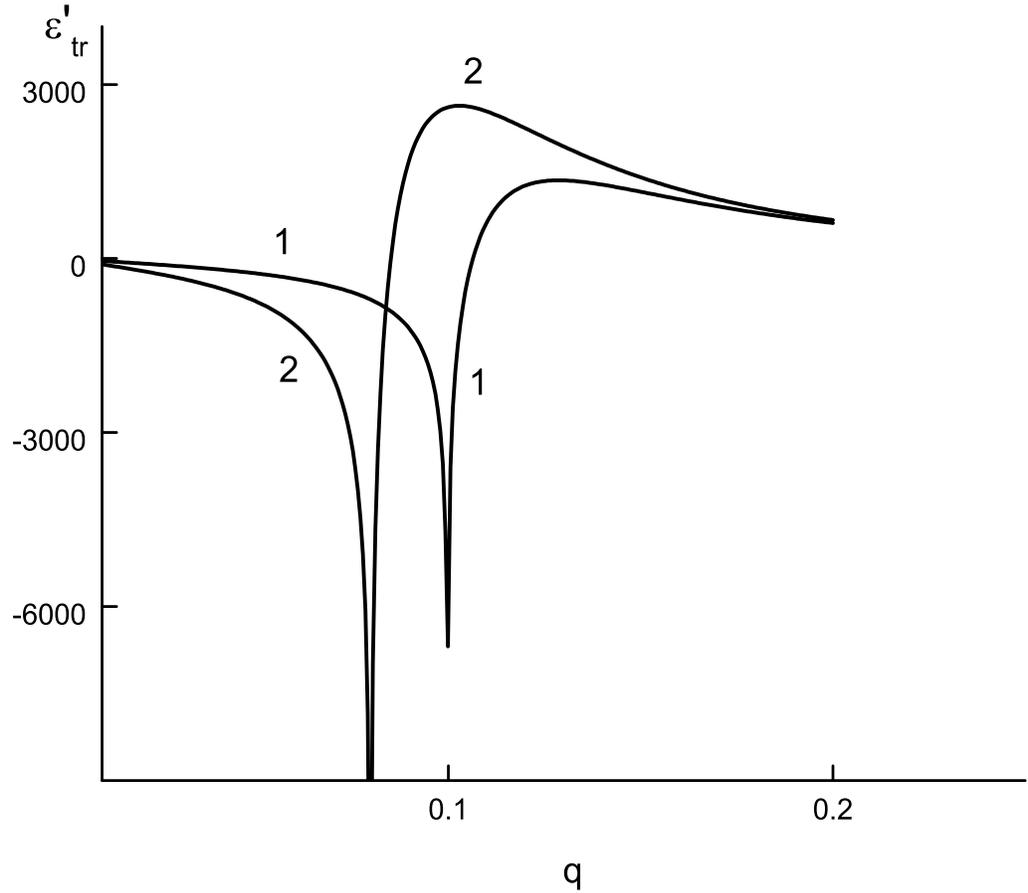}
\caption{Kohn kind singularities: derivative of
the dielectric permittivity, curves $1$ and $2$ correspond
to the values $\Omega=0.1$ and
$\Omega=0.08$.
}\label{rateIII}
\end{flushleft}
\end{figure}

Let us carry out the change of varible of integration
$k_1=q/\varepsilon$ and note that
$$
k_1x_1=k_1\varepsilon\dfrac{x_1}{\varepsilon}=q\dfrac{x}{l
\varepsilon}=q\dfrac{x\omega_p}{v_F\tau \nu}=q\dfrac{\omega_p}{v_F}x.
$$

Then
$$
\dfrac{E(x)}{E'(0)}=
\dfrac{al}{\varepsilon\pi}\int\limits_{-\infty}^{\infty}
\dfrac{e^{iq\omega_p x/v_F}dq}{\varepsilon_{tr}(q)-bq^2},
$$
where
$$
b=\dfrac{a}{\varepsilon^2}=\Big(\dfrac{c}{v_F\Omega}\Big)^2.
$$

Our aim is to analyze the asymptotic behaviour of the electric field
under $x\to\infty$. At the same time we would consider the
contribution into the integral of region near of the derivative
singularity $\varepsilon_{tr}$. Our consideration would be similar
to the one stated into \cite{Harrison}.

Integrating twice by parts we receive
$$
\dfrac{E(x)}{E'(0)}=\dfrac{alv_F^2}{\varepsilon \pi \omega_p^2 x^2}
\int\limits_{-\infty}^{\infty}
\dfrac{\varepsilon_{tr}''(q)e^{iq\omega_px/v_F}\,dq}
{[\varepsilon_{tr}(q)-bq^2]^2}.
\eqno{(3)}
$$

In the expression (3) we write down only terms which display the most
anomalous behaviour near the Kohn singularity. In this approximation
 $\varepsilon''_{tr}(q)$ we can write in the following form
$$
\varepsilon_{tr}''(q)=-\dfrac{3}{4\Omega q^3}
\Big[\dfrac{\Omega+i\varepsilon+q}{\Omega+i\varepsilon-q}-
\dfrac{\Omega+i\varepsilon-q}{\Omega+i\varepsilon+q}\Big].
$$

Now instead of (3) for the electric field we obtain the following
expression
$$
\dfrac{E(x)}{E'(0)}=\dfrac{3c^2v_F}{4\pi \omega^2 \omega_p x^2}
\int\limits_{-\infty}^{\infty}
\Big[\dfrac{q+\Omega+i\varepsilon}{q-\Omega-i\varepsilon}-
\dfrac{q-\Omega-i\varepsilon}{q+\Omega+i\varepsilon}\Big]
\dfrac{e^{iq \omega_p x/v_F}\,dq}{q^3[\varepsilon_{tr}(q)-bq^2]^2}.
\eqno{(4)}
$$

Let us consider the case of colissionless plasma, i.e. the case when
$\varepsilon\to 0$. Then the integral (4) can be simplified
significantly:
$$
\dfrac{E(x)}{E'(0)}=\dfrac{3c^2v_F}{4\pi \omega^2 \omega_p x^2}
\int\limits_{-\infty}^{\infty}
\Big[\dfrac{q+\Omega}{q-\Omega}-
\dfrac{q-\Omega}{q+\Omega}\Big]
\dfrac{e^{iq \omega_p x/v_F}\,dq}{q^3[\varepsilon_{tr}(q)-bq^2]^2}.
\eqno{(5)}
$$

Here
$$
\varepsilon_{tr}(q)-bq^2=1-\dfrac{3}{2q^2}-\dfrac{3}{4\Omega q^3}
(\Omega^2-q^2)\ln\dfrac{\Omega-q}{\Omega+q}-
\Big(\dfrac{cq}{v_F\Omega}\Big)^2.
$$

For the calculation of the integral from (5)
$$
J=\int\limits_{-\infty}^{\infty}
\Big[\dfrac{q+\Omega}{q-\Omega}-
\dfrac{q-\Omega}{q+\Omega}\Big]
\dfrac{e^{iq \omega_p x/v_F}\,dq}{q^3[\varepsilon_{tr}(q)-bq^2]^2}
$$
we use the method stated in the monograph \cite{Harrison}. For this
purpose we would present the integral $J$ in the form of the
difference $J=J_1-J_2$, where
$$
J_1=\int\limits_{-\infty}^{\infty}
\dfrac{q+\Omega}{q-\Omega}\cdot
\dfrac{e^{iq \omega_p x/v_F}\,dq}{q^3[\varepsilon_{tr}(q)-bq^2]^2},
$$
$$
J_2=\int\limits_{-\infty}^{\infty}
\dfrac{q-\Omega}{q+\Omega}\cdot
\dfrac{e^{iq \omega_p x/v_F}\,dq}{q^3[\varepsilon_{tr}(q)-bq^2]^2},
$$

After the evident change of variable of integration for the second
integral  $J_2$ we get the expression:
$$
J_2=-\int\limits_{-\infty}^{\infty}
\dfrac{q+\Omega}{q-\Omega}\cdot
\dfrac{e^{-iq \omega_p x/v_F}\,dq}{q^3[\varepsilon_{tr}(q)-bq^2]^2}.
$$

Consequently, the integral $J$ equals to:
$$
J=2\int\limits_{-\infty}^{\infty}
\dfrac{q+\Omega}{q-\Omega}\cdot
\dfrac{\cos\Big(\dfrac{q \omega_p x}{v_F}\Big)}
{q^3[\varepsilon_{tr}(q)-bq^2]^2}dq.
$$

Considering the singularity of the kernel $1/(q-\Omega)$ of this
integral the most large contribution to the value of this integral
is made by the values of the subintegral function near the point of
singularity  $q=\Omega$. The function
$f(q)=(q+\Omega)q^{-3}[\varepsilon_{tr}(q)-bq^2]^{-2}$ change slowly
in the neighbourhood of the point $q=\Omega$. Therefore further we
assume $f(q)=f( \Omega)$ under the calculation of the integral $J$
near the singular point. We obtain
$$
J=2 f(\Omega)\int\limits_{-\infty}^{\infty}
\dfrac{\cos\Big(\dfrac{q \omega_p x}{v_F}\Big)}
{q-\Omega}dq.
$$

After the integration variable change $q-\Omega=u$,  noticing that
$$
\cos\Big(\dfrac{q \omega_p x}{v_F}\Big)=
\cos\Big(\dfrac{(u+\Omega) \omega_p x}{v_F}\Big)=
\cos \dfrac{\omega_p u}{v_F}x\cos \dfrac{\omega x}{v_F}-
\sin \dfrac{\omega x}{v_F}\sin\dfrac{\omega_p u}{v_F}x,
$$
and using the known relation
$$
\int\limits_{-\infty}^{\infty}\dfrac{\sin Ax}{x}dx=\pi,
$$
we receive:
$$
J=-2\pi f(\Omega)\sin \dfrac{\omega}{v_F}x,
$$
where
$$
f(\Omega)=\dfrac{2}{\Omega^2\Big[\dfrac{3}{2\Omega^2}+
\Big(\dfrac{c}{v_F}\Big)^2-1\Big]^2}=
\dfrac{2\omega^2\omega_p^2}{\Big[\dfrac{3}{2}\omega_p^2+
\Big(\dfrac{c}{v_F}\Big)^2\omega^2-\omega^2\Big]^2},
$$
or
$$
f(\Omega)=\dfrac{2\Omega^2}{\Big[\dfrac{3}{2}+
\Big(\dfrac{c}{v_F}\Omega\Big)^2-\Omega^2\Big]^2}.
$$

Thereby we have found that the electric field far away from the
surface $x=0$ decreases according to the law:
$$
\dfrac{E(x)}{E'(0)}=-\dfrac{A}{x^2}\sin \dfrac{\omega}{v_F}x,
\eqno{(6)}
$$
where
$$
A=\dfrac{3c^2v_F}{x^2\omega_p^3}\cdot \dfrac{1}
{\Big[\dfrac{3}{2}+\Big(\dfrac{c\Omega}{v_F}\Big)^2-\Omega^2\Big]^2}=
\dfrac{3c^2v_F\omega_p}{\Big[
\dfrac{3\omega_p^2}{2}+\dfrac{c^2}{v_F^2}\omega^2-\omega^2\Big]^2}.
$$

Let us rewrite the formula (6) with the help of the dimensionless
parameters in the following form:
$$
\dfrac{E(x)}{E'(0)}=-\dfrac{A}{x^2}\sin \dfrac{\Omega \omega_p}{v_F}x,
\eqno{(7)}
$$
where
$$
A=\dfrac{3c^2v_F}{\omega_p^3\Big[\Omega^2\Big(\dfrac{c^2}{v_F^2}-1\Big)+
\dfrac{3}{2}\Big]^2}.
\eqno{(8)}
$$

Noting that in the considered here nonrelativistic case
 $v_F\ll c$, we can simplify the formula (8):
$$
A=\dfrac{3c^2v_F}{\omega_p^3\Big[\dfrac{3}{2}+
\Big(\dfrac{c}{v_F}\Omega\Big)^2\Big]^2}.
\eqno{(9)}
$$

In the low frequencies case when $\Omega\ll \dfrac{v_F}{c}$, the
formula (9) can be simplified and reduced to the following form:
$$
A=\dfrac{4c^2v_F}{3\omega_p^3},
$$
and in the case when $\Omega\gg \dfrac{v_F}{c}$, the formula (9)
can be reduced to the form
$$
A=\dfrac{3v_F^5}{c^2\omega_p^3\Omega^4}.
$$

Let us consider the case of infra-red frequencies. The formula (1) is
applicable near the surface. In accordance with this fact we obtain
$$
\dfrac{E(0)}{E'(0)}=-\dfrac{c}{\omega_p}.
\eqno{(10)}
$$

Dividing the equality (7) by (10), we have:
$$
E(x)=\dfrac{\omega_p}{c}\dfrac{A(\Omega,v_F)E(0)}{x^2}
\sin\Big(\dfrac{\omega_p\Omega}{v_F}x\Big).
\eqno{(11)}
$$

The formula (11) can be presented in the form
\[
E(x)=\dfrac{B}{x^2}\sin\Big(\dfrac{\omega_p\Omega}{v_F}x\Big)E(0),\qquad
B=\dfrac{3cv_FE(0)}{\omega_p^2\Big[\dfrac{3}{2}+
\Big(\dfrac{c}{v_F}\Omega\Big)^2-\Omega^2\Big]^2}.
\]

Let us carry out the graphic research of the expressions obtained.
The behaviour of the variable $y=|E(x)/E(0)|$ is shown on the Figs.
2--4, where the relation $E(x)/E(0)$ is determined according to the
equality (11). The distance on the horizontal axis is measured in
centimetres.

We present the behaviour of the curves $y_1(x)=B/x^2\;(E(0)=1)$ (curve $1$)
and $y_2(x)=e^{-\omega_px/c}$ (curve $2$) on the Figs. 5 and 6. At
the same time the logarithmic scale is used on the vertical axis,
and the distance on the horizontal axis is measured in microns.

\begin{center}\bf
3. Conclusion
\end{center}

Friedel was the first \cite{friedel1} -- \cite{friedel5} to discover
that asymptotic (on the large distances) decreasing of the screened
potential of the point charge under quantum consideration of the
degenerate collisionless plasma has not only monotonously decreasing
but also oscillating character as well. The reason of such
oscillations is the sharp falling (to the zero) beyond the Fermi
surface of the Fermi distribution for the electrons  $f_F(v)$,
$$
f_F(v)=\Theta(v_F-v),
$$
$\Theta(x)$ is the Heaviside function,
$$
\Theta(x)= \left\{ \begin{array}{c}
                    1, \quad x>0, \\
                    0, \quad x<0.
                  \end{array}\right.
$$

This peculiarity of the Fermi distribution results in so called Konh
singularities   (see \cite{Kohn0} -- \cite{Grassme}). Kohn
singularities are the consequences of the logarithmic singularities
of the longitudinal dielectric permittivity of the degenerate
plasma. Just the Kohn singularities result in the Friedel
oscillations.

In the work \cite{Grassme} the dependance of the Friedel
oscillations on the tempe\-ra\-tu\-re in the collisionless plasma is found
out. It was shown that under finite temperature the amplitude of the
Friedel oscillation decreases exponen\-ti\-ally with the distance.
Similar dependance should be apparent under Friedel oscillations in
skin effect as well.

\begin{figure}[h]
\includegraphics[width=16.0cm, height=10cm]{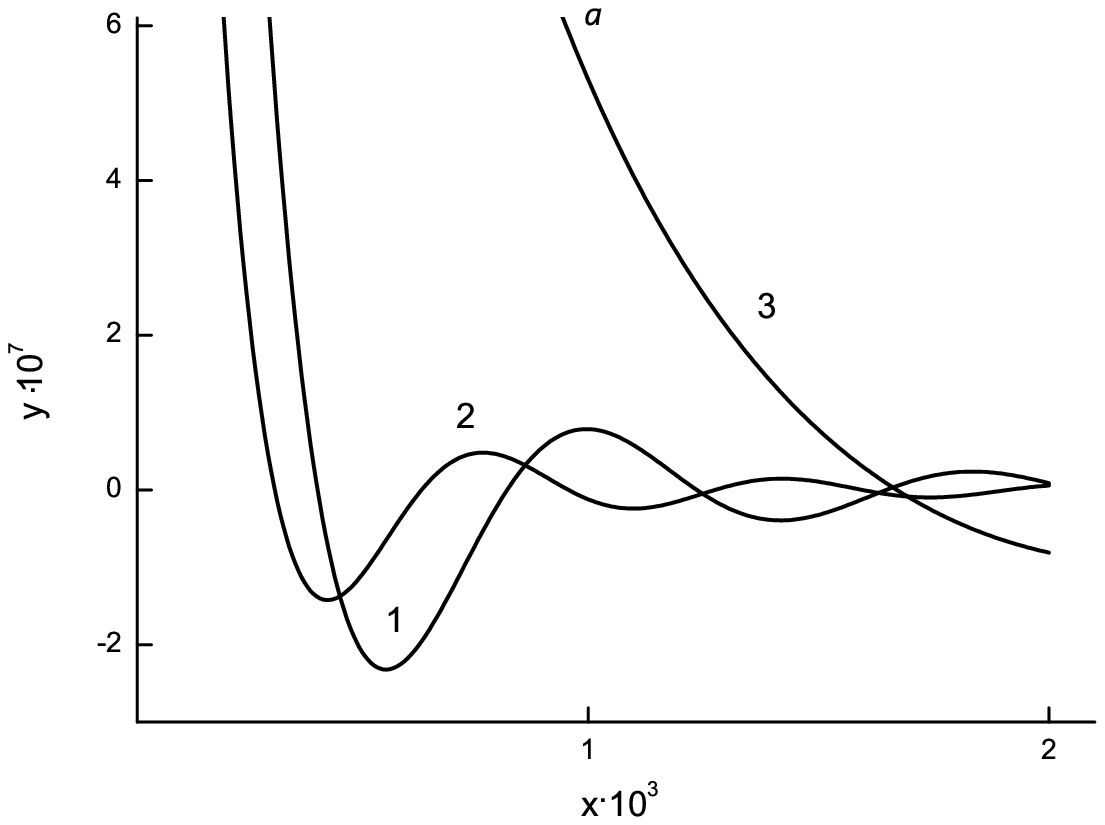}
\caption{Oscillations of kind of Friedel in the case
$\Omega=10^{-4}$, $2\cdot10^{-5}<x<2\cdot 10^{-3}$. Curves
$1,2,3$ correspond to sodium, gold and aluminium.}
\end{figure}

\begin{figure}[h]
\includegraphics[width=16.0cm, height=9.5cm]{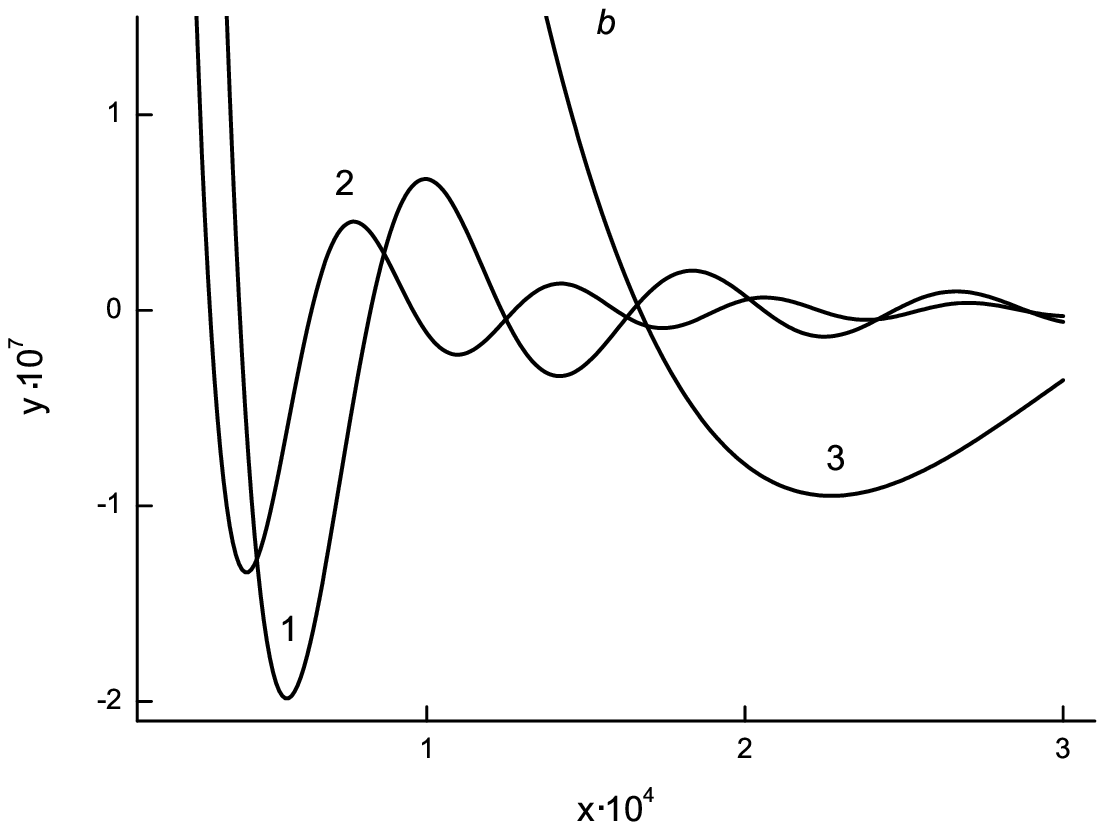}
\caption{Oscillations of kind of Friedel in the case $\Omega=10^{-3}$,
$9\cdot 10^{-5}<x<3\cdot 10^{-3}$. Curves
$1,2,3$ correspond to sodium, gold and aluminium.}
\end{figure}

\begin{figure}[h]
\begin{flushleft}
\includegraphics[width=16.0cm, height=9.5cm]{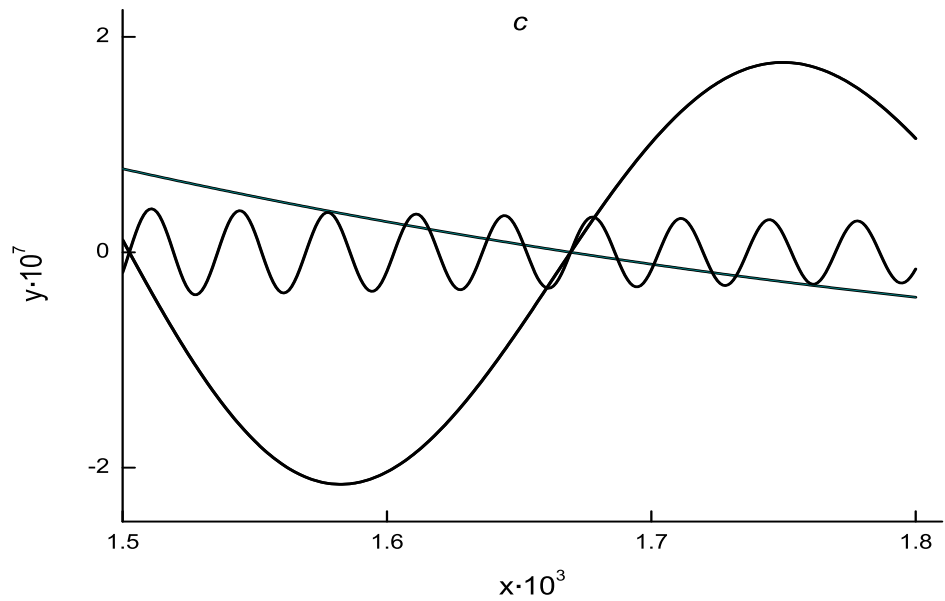}
\caption{Oscillations of kind of Friedel in the case of aluminium,
 $1.5\cdot 10^{-3}<x<1.8\cdot 10^{-3}$. Curves $1,2,3$ correspond
 to the values $\Omega=10^{-4}, 10^{-3}, 10^{-2}$.
}\label{rateIII}
\end{flushleft}
\end{figure}

\begin{figure}[h]
\begin{flushleft}
\includegraphics[width=16.0cm, height=10cm]{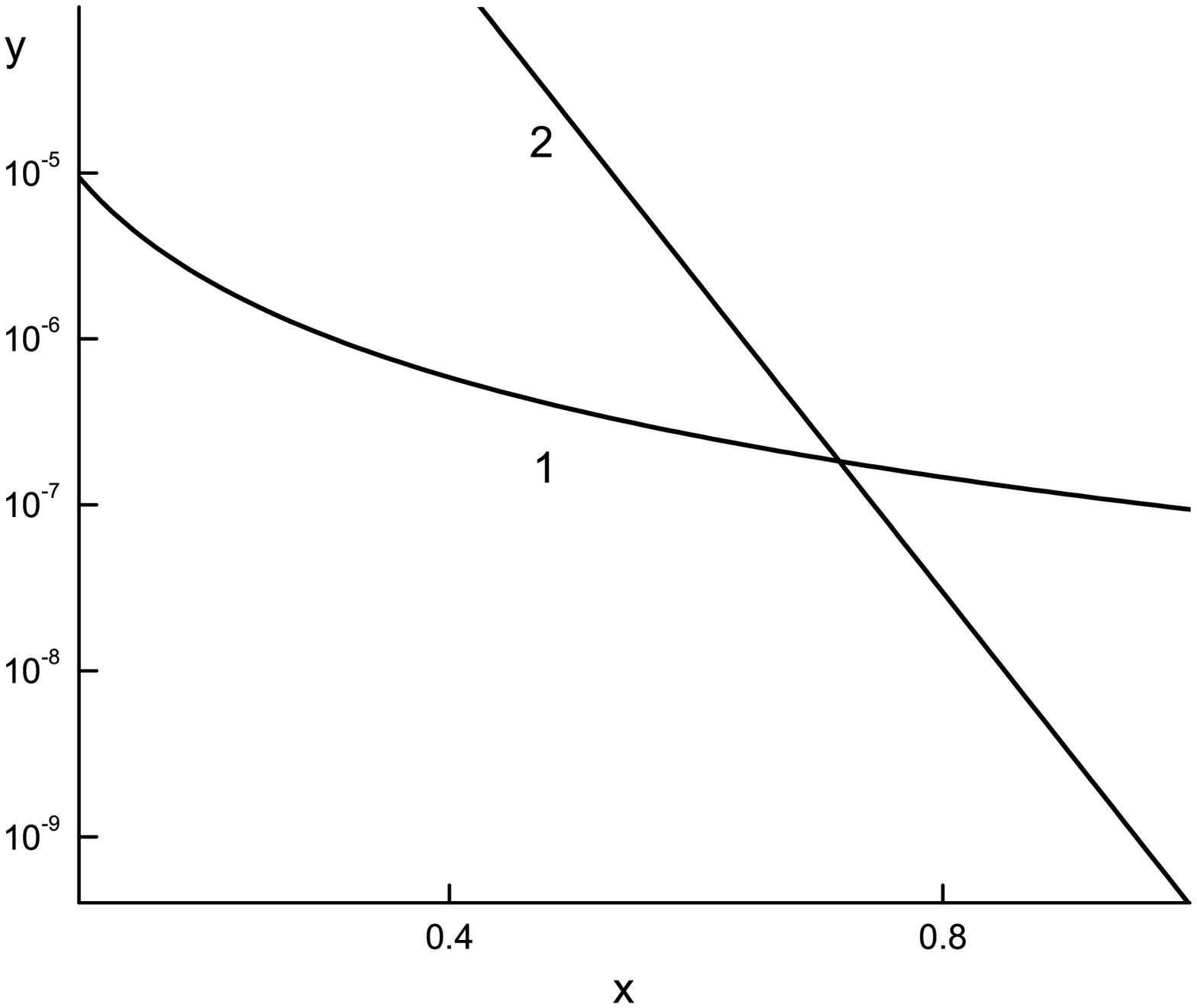}
\caption{Intersection of the two curves $y_1(x)=B/x^2$ (curve $1$)
and $y_2(x)=
e^{-\omega_px/c}$ (curve $2$) in the point $x_*=0.716$
micrometers under $\Omega=10^{-2}$
(logarithmic scale on the vertical axis).
}\label{rateIII}
\end{flushleft}
\begin{flushleft}
\includegraphics[width=16.0cm, height=9cm]{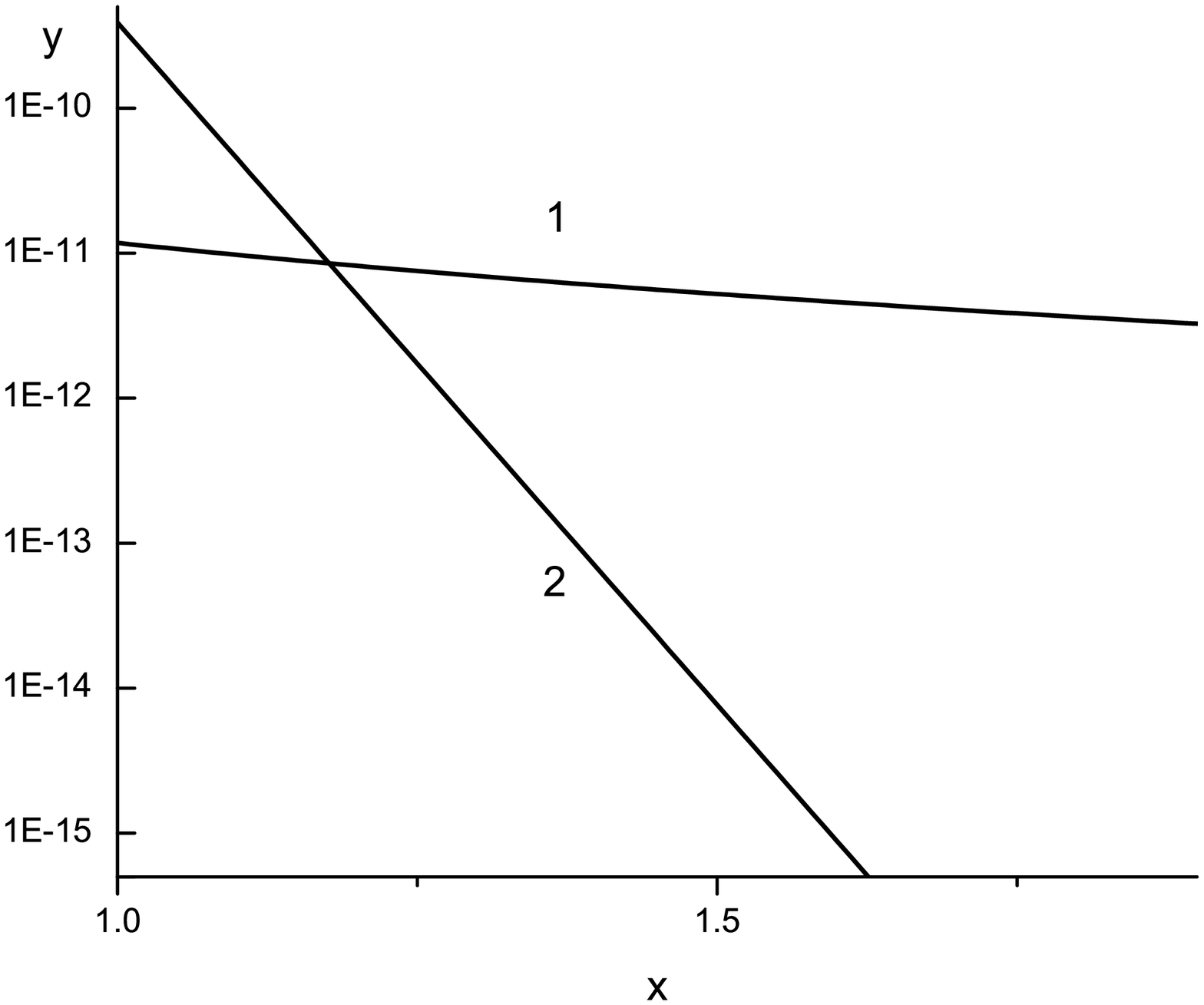}
\caption{Intersection of the two curves $y_1(x)=B/x^2$
(curve $1$) and $y_2(x)=
e^{-\omega_px/c}$ (curve $2$) in the point $x_*=1.176$
micrometers under $\Omega=10^{-1}$
(logarithmic scale on the vertical axis).
}
\end{flushleft}
\end{figure}

\end{document}